# A Low-Code Methodology for Developing AI Kiosks: a Case Study with the DIZEST Platform

SunMin Moon
*S/W Development Center*
SEASON Co. Ltd.
Sejong, Korea
judians@season.co.kr

Jangwon Gim
*Dept. of Software*
Kunsan National University
Gunsan, Korea
jwgim@kunsan.ac.kr

Chaerin Kim
*R&D Center*
Jenti AI
Seoul, Korea
cherry@jenti.ai

Yeeun Kim
*R&D Center*
SEASON Co. Ltd.
Sejong, Korea
loyevei@season.co.kr

YoungJoo Kim
*S/W Development Center*
SEASON Co. Ltd.
Sejong, Korea
jootopia@season.co.kr

Kang Choi*
*R&D Center*
SEASON Co. Ltd.
Sejong, Korea
river@season.co.kr

*Corresponding: river@season.co.kr

*Abstract*—This paper presents a comprehensive study on enhancing kiosk systems through a low-code architecture, with a focus on AI-based implementations. Modern kiosk systems are confronted with significant challenges, including a lack of integration, structural rigidity, performance bottlenecks, and the absence of collaborative frameworks. To overcome these limitations, we propose a DIZEST-based approach methodology, a specialized low-code platform that enables intuitive workflow design and seamless AI integration. Through a comparative analysis with existing platforms, including Jupyter Notebook, ComfyUI, and Orange3, we demonstrate that DIZEST delivers superior performance across key evaluation criteria. Our photo kiosk case study further validates the effectiveness of this approach in improving interoperability, enhancing user experience, and increasing deployment flexibility.

## I. Introduction

In modern society, kiosks are widely utilized in various forms. Kiosk systems are spotlighted as a means to reduce labor burdens and enhance service efficiency, and their market size continues to grow steadily [1]. Recently, by integrating cutting-edge technologies such as AI, IoT, and machine learning, kiosks have evolved into more personalized and intelligent services, continuously expanding their range of applications and raising user expectations.

However, despite these technological advances and market growth, modern kiosk systems still face several common limitations and challenges. First, kiosks individually designed by different manufacturers and for various purposes often lack backend system integration and interoperability [2][3], making data linkage and service expansion difficult. Second, existing kiosk software is primarily built using monolithic architectures, resulting in insufficient flexibility and modularity, which restricts the addition of new features or customized deployments. Third, performance degradation on outdated hardware or under-optimized systems deteriorates the user experience and can ultimately lead to service abandonment [4]. Fourth, the absence of collaborative support structures makes it difficult for multiple developers and domain experts to develop or modify systems simultaneously and rapidly, hindering continuous improvement and quality management. These limitations—including lack of integration, structural rigidity, performance bottlenecks, and absence of collaboration—are recognized as significant challenges in current kiosk systems. Fifth, in real service environments, multiple users often interact with kiosks simultaneously, which introduces challenges in real-time user tracking [5]. Reliable user re-identification also becomes difficult under occlusion, making it a critical issue in practical deployments [6].

This paper seeks to address the limitations faced by kiosk systems, utilizing photo kiosks as a specific case study. As photo kiosks provide multimedia services such as photo printing and editing, they serve as a representative example for comprehensively examining and addressing core issues such as interoperability, UX accessibility, and system flexibility. The importance of user-friendly interfaces in kiosk systems has been well- documented in previous research [7], particularly regarding age-specific UI enhancements that improve overall system usability. Based on this analysis, this study proposes the application of our proprietary low-code platform, DIZEST, to photo kiosks to improve interoperability, enhance user experience, increase maintenance efficiency, and boost deployment flexibility.

DIZEST is a technical solution designed with a modular architecture and cloud connectivity, which facilitates easy

This work was supported by the Korea Technology and Information Promotion Agency for SMEs(TIPA) grant funded by the Korea government (grant number: RS-2024-00446996).

development and deployment of kiosk services and allows seamless integration of AI-based intelligent features. Through this research, we validate the effectiveness of applying the DIZEST platform to address the aforementioned limitations and present practical solutions to overcome the challenges faced by modern kiosk systems. Furthermore, we suggest a methodology for developing a versatile low-code platform that can be applied to various domains.

## II. Requirements for Modern Kiosk Systems

### A. R1: Configuration modification through intuitive UI/UX (Visual Accessibility)

Kiosk systems should provide an intuitive UI/UX that enables non-expert administrators to easily modify settings on-site [4][7]. This is especially important in situations where immediate response and adjustments are required. Conventional administrator pages typically offer only pre-defined functions, making it challenging to handle unexpected scenarios or highly customized requirements flexibly. Such a flexible structure is essential for achieving both rapid development of tailored services and operational efficiency in diverse service environments. Furthermore, continuous feature improvement and service enhancement are supported, which are critical for realizing a flexible system architecture. It is also necessary to improve the visual accessibility and comprehensibility of the UI to ensure that administrators can operate the system effortlessly.

### B. R2: Separate management system for ensuring system stability

Kiosk systems have a complex structure that integrates various functions to provide services to users. Therefore, even when a specific function fails, users should be able to recognize the error while continuing to use the core services. Additionally, system errors or abnormal behaviors should be detected and handled in an independent environment separated from the operational environment. Such separated management and error-handling mechanisms are essential for maintaining system robustness and reliability, minimizing potential service disruptions, and ultimately securing high availability [3].

### C. R3: Streamlined deployment and minimization of system complexity

The software of kiosk systems must be lightweight, and the complexity of deployment and system configuration should be minimized. This consideration is especially important in resource-constrained environments or when building large-scale kiosk networks [3]. If the software becomes heavy, it can excessively consume system resources and slow down processing speed, which negatively affects the overall service stability. Additionally, complicated installation processes or heavy system requirements increase initial deployment costs and make maintenance more difficult. Therefore, a simplified deployment procedure and an architecture that enables efficient resource utilization are essential so that users can easily install and manage the system. This approach helps reduce maintenance costs, improve service reliability and response speed, and ultimately maximize operational efficiency.

### D. R4: Modularity and reusability

In the process of implementing actual kiosk systems, we have identified the necessity of modularizing AI-based functions to ensure flexible utilization [2]. The AI-related features in kiosk systems, such as background or costume transformations, must be modularized so that they can be organically utilized depending on the situation. This means that each function operates as an independent component that can be combined or replaced as needed. Such modularity enhances reusability, facilitates updates and improvements of AI models, and strengthens development and maintenance efficiency. This structure supports rapid development of customized services and operational efficiency in various service environments. Furthermore, it is essential for realizing a flexible system architecture that can quickly respond to diverse operational environments while supporting continuous feature improvements and service advancements.

## III. Implementation of a Photo Kiosk System Using DIZEST

### A. System Architecture for AI-Based Kiosk

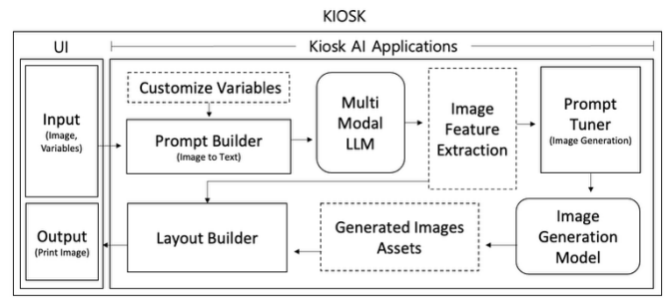

Fig. 1. Overall architecture of the AI-based kiosk system.

The photo kiosk system proposed in this study provides a service that generates and delivers AI-based images and outputs in real time, based on user-input face images, tailored to different historical eras and concepts.

Figure 1 illustrates the overall architecture of the proposed AI kiosk system, which is designed to provide real-time personalized image generation and printing services. The system consists of two main layers: the User Interface (UI) and Kiosk AI Applications. The UI layer includes the Input module, which receives user-provided images and customization variables, and the Output module, which delivers the final print-ready images. Within the AI applications layer, the

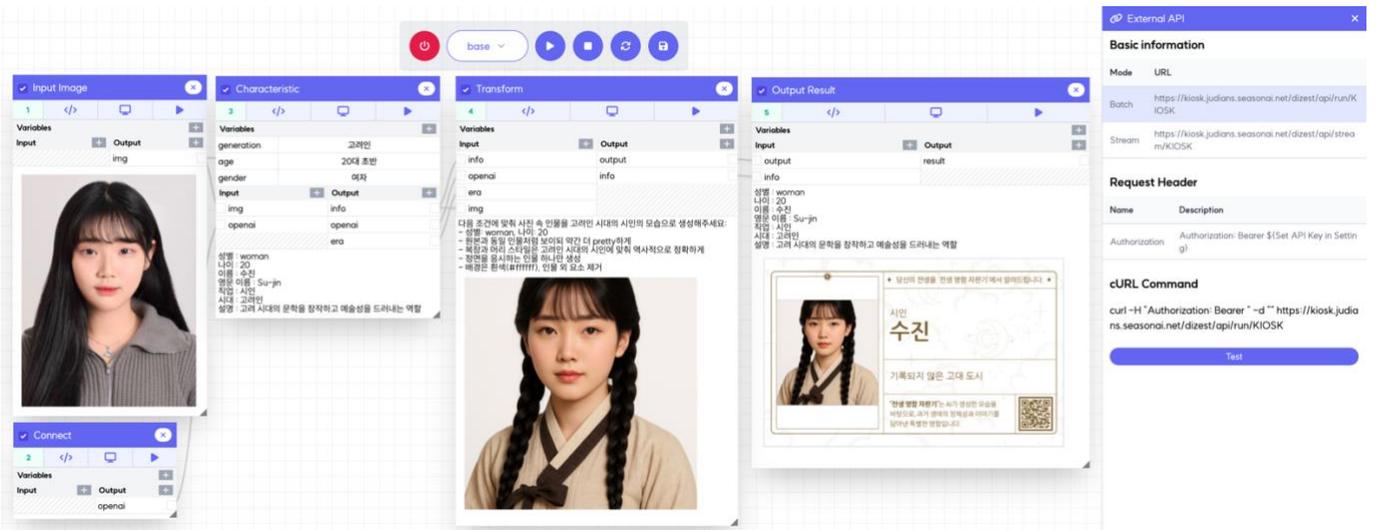

Fig. 2. DIZEST workflow layout for the proposed kiosk system.

Prompt Builder converts input images and variables into descriptive textual prompts (image-to-text), which can be further adjusted through the Customize Variables function to reflect user preferences more precisely. The prompts are then processed by the Multi-Modal Large Language Model (LLM), which performs high-level semantic analysis and integrates visual and textual contexts. The Image Feature Extraction module extracts detailed visual attributes that guide the generation process.

Subsequently, the Prompt Tuner refines and optimizes the prompts specifically for the image generation task. The Image Generation Model uses these optimized prompts to synthesize personalized images tailored to the user's intended style or historical concept. Generated images are stored as intermediate Generated Image Assets and then composed into a final layout through the Layout Builder, ensuring that the output is visually coherent and suitable for immediate printing. This modular architecture allows even non-expert users to easily interact with and control the end-to-end image generation and printing process while maintaining high-quality, customizable outputs.

*B. Proposed implementation methodology*

To realize this system, we utilized the DIZEST platform, which combines intuitive workflow design with AI image transformation technology [8]. This platform provides a canvas-based visual workflow design environment, where modular functional blocks (Apps) can be combined to visually configure the entire pipeline—from image processing and feature extraction to AI model transformation and final result output. Figure 2 shows this platform, illustrating how users can intuitively design and control each step of the kiosk system process through a DIZEST-style visual interface. To effectively address diverse requirements inherent in real-world kiosk deployments, DIZEST's technical solutions were comprehensively applied, leveraging its core capabilities in several key aspects:

First, the complexity of AI is visualized through simple diagrams, supporting intuitive UI/UX-based configuration modifications. As a result, even non-expert users can easily construct and modify AI workflows on a web-based canvas via drag-and-drop, and adjust variables on-site without source code changes, thereby improving accessibility and usability.

Second, the kiosk environment and AI execution environment are designed as separate structures to ensure system stability. AI workflows are executed on the DIZEST server, while kiosks call the server's API to use the functionalities. This approach isolates AI processing loads and errors from affecting the kiosk, enhancing overall system stability.

Third, DIZEST can be easily deployed via Python packages and Docker images, and it simplifies external system integration through REST APIs. This architecture streamlines integration processes, reduces system weight, and contributes to operational cost savings.

Finally, AI functions are modularized into "Apps" to enhance modularity and reusability. Each App operates as a functional block that can be recombined and expanded within various workflows. Based on the OSMU (One Source Multi Use) concept, the same modules can be repeatedly reused across different scenarios, improving development efficiency and scalability.

IV. COMPARATIVE ANALYSIS WITH EXISTING PLATFORMS

The AI photo kiosk system must meet four key requirements—on-site immediacy, operational stability,

functional flexibility, and lightweight deployment—in order to secure practical service value and scalability. Towards this goal, this study conducted a comparative analysis of Jupyter Notebook, ComfyUI, Orange3, and the proposed DIZEST platform.

TABLE I. COMPARISON BETWEEN EXISTING PLATFORMS AND DIZEST

| Platforms | Requirements | | | | |
| --- | --- | --- | --- | --- | --- |
| | *R1* | *R2* | *R3* | *R4* | *Reference* |
| Jupyter | L | L | M | L | [9] |
| ComfyUI | M | L | M | L | [10, 11] |
| Orange3 | M | M | M | M | [12] |
| DIZEST | H | H | H | H | [13] |

a. High / Medium / Low

Table 1 presents a comparative analysis of Jupyter, ComfyUI, Orange3, and the proposed DIZEST platform based on key requirements. While the existing platforms each have their unique advantages, DIZEST received high scores across all evaluation criteria and demonstrated superior performance, particularly in on-site applicability and scalability. A detailed examination of each existing platform's characteristics, including their respective strengths and limitations, follows to further contextualize these findings.

TABLE II. PLANNED QUANTITATIVE EVALUATION METRICS FOR DIZEST

| *Domain* | *Metrics* | *Measurement Plan* |
| --- | --- | --- |
| Development Efficiency | Workflow design and deployment time savings (%) | Record time compared to traditional methods |
| Operational Stability | System uptime (%), Mean Time to Recovery (MTTR) | Monitor logs and measure recovery time during failures |
| Cost Efficiency | Maintenance costm reduction, deployment/operation overhead reduction | Analyze cost data before and after deployment |
| User Experience | Administrator and user satisfaction (Likert scale) | Conduct surveys and perform statistical analysis |

Jupyter Notebook provides a code cell-based development environment that is widely used by data scientists and developers. While it offers a high degree of freedom and flexibility, it has the limitation of a steep learning curve for non-experts. Due to its architecture, where the kernel and UI are tightly coupled, an error during execution can potentially halt the entire system. Although function-level reuse is possible, overall maintenance and deployment are complex. In particular, the complexity of notebook duplication and deployment processes hinders collaboration and scalability.

ComfyUI offers a node-based UI, enabling visual workflow construction, which is a major advantage. However, there remains complexity in variable modifications, and since it operates based on local GPUs, it is difficult to separate the AI execution environment from the UI environment. This introduces vulnerabilities in terms of system stability. While node reuse is possible, there are limitations in reconfigurability, restricting scalability to diverse scenarios. Moreover, its reliance on local installations and GPU dependencies makes lightweight deployment and integration with external systems challenging.

Orange3 also supports data mining and visualization through a simple, node-based UI. However, it presents significant limitations for enterprise-level applications. Its local-centric environment makes implementing complex logic and managing large systems difficult. Consequently, despite its basic modularity and simple installation, Orange3's limited scalability poses a major challenge for integration and expansion in an enterprise setting.

In contrast to these existing solutions, the proposed DIZEST platform is designed to overcome the aforementioned limitations, offering robust performance across all evaluation criteria. As evidenced in Table 1, DIZEST consistently achieved high scores in on-site immediacy, operational stability, functional flexibility, and lightweight deployment. Its architecture specifically addresses the challenges of tightly coupled components, limited scalability, and complex deployment processes inherent in other platforms, thereby ensuring superior on-site applicability and comprehensive scalability for diverse service environments.

## V. CONCLUSION

This study proposed the DIZEST platform to address the key limitations faced by modern kiosk systems, including lack of integration, structural rigidity, performance bottlenecks, and absence of collaborative frameworks. Based on empirical analysis using a photo kiosk case, we derived four essential requirements for next-generation kiosk systems: intuitive UI/UX-based configuration modification, a separate management system for system stability, modularization and reusability of AI functions, and lightweight deployment with minimized system complexity.

The DIZEST platform enables even non-expert users to intuitively design and operate kiosk services through a canvas-based drag-and-drop interface. It ensures system stability via a distributed architecture, maximizes AI function reusability through App-level modularization, and enhances operational efficiency through lightweight deployment using Container-based architecture.

The comparative analysis with other platforms demonstrated that DIZEST delivers superior performance in terms of accessibility for on-site operators, system stability, module reusability, and ease of deployment.

In summary, the DIZEST platform is expected to overcome the current limitations of modern kiosk systems and contribute to the digital transformation and service advancement of the kiosk industry by providing a scalable solution adaptable to various kiosk environments.

As a follow-up study, we plan to quantitatively evaluate key performance indicators—such as reduced development time, lower maintenance costs, increased system uptime, and improved end-user satisfaction—during actual commercial deployments of DIZEST. Specifically, this evaluation will measure workflow design and deployment time savings compared to traditional environments, track system uptime and mean time to recovery (MTTR), analyze reductions in maintenance and operational overhead costs, and collect administrator and end-user satisfaction data through Likert-scale surveys. These efforts aim to validate its practical business value and further enhance the platform's efficiency.